\documentclass[12pt]{article}
\usepackage{amssymb,amsmath,amsthm,amsfonts,amscd}
 \usepackage{graphicx}
\newcommand\be{\begin{equation}}
\newcommand\ee{\end{equation}}
\newcommand\bea{\begin{eqnarray}}
\newcommand\eea{\end{eqnarray}}

\newcommand{\fatalpha}{{\bf \alpha \kern -0.44em \alpha}}
\newcommand{\fatsigma}{{\bf \sigma \kern -0.54em \sigma}}
\newcommand{\tpchi}{{\bf D \kern -0.35em D}}
\newcommand{\llambda}{{\bf \lambda \kern -0.45em \lambda}}

\bibliography{plain}
\title{\bf Kraus representation for maps and master equation in spin star model with layered environment} \vspace{20mm}
\author{ M.Mahdian$^{a}$
 \thanks{E-mail:Mahdian@tabrizu.ac.ir}
 , H. Mehrabpour$^{a}$ ,
 \\
$^a${\small Department of Theoretical Physics and Astrophysics,
University of Tabriz, Tabriz 51664, Iran.}  } \pagebreak

\vspace{20mm}

\begin{document}
\maketitle \vspace{15mm}
\newpage
\begin{abstract}
Quantum operations are usually defined as completely positive (CP),
trace preserving (TP) maps on quantum states, and can be represented
by operator-sum or Kraus representations. In this paper, we
calculate operator-sum representation and master equation of an
exactly solvable dynamic of one-qubit open system in layered
environment . On the other hand, we obtain exact Nakajima-Zwanzig
(NZ) and time-convolutionless (TCL) master equation from the maps.
Finally, we study a simple example to consider the relation between
CP maps and initial quantum correlation and show that vanishing
initial quantum correlation is not necessary for CP maps.


\end{abstract}
\section{Introduction}  
In quantum information theory, it is widely accepted that any
physical process can be described by a quantum operation or quantum
channel which is CP (this means that not only  map $\Phi:
B_{1}\rightarrow B_{2}$ on $c^{\star}$-algebra is positive, but also
the combined operation $\Phi\otimes I_N:B_{1}\otimes N\rightarrow
B_{2}\otimes N$ for all dimensions N is positivity preserving ) and
TP maps (this means that $\mathrm{Tr}\Phi(\rho)=\mathrm{Tr} \rho$
for all trace class operator $\rho$) between spaces of
operators\cite{MKeyl}-\cite{PWShor}.  These maps play an important
role in the description of nonunitary time evolutions of open
quantum systems that interact with an environment. On the other
hand, the environment is classified as Markovian with no memory
effect and non-Markovian  with memory effect
\cite{RVasile}-\cite{MANielsen}.

It is well-known that for closed quantum system, its time evolution
can be described by a unitary operator. However, for an open system,
the time evolution is not necessarily unitary. The evolution of an
open system is usually described by the Kraus representation
\cite{KKraus}. The Kraus representation of an open system is usually
constructed by considering a large closed system. Let's assume an
interaction between the open system denoted as ($s$) and the
environment ($b$). This environment is a quantum system with the
Hilbert space of an arbitrary dimension. The whole ($s$)+($b$)
system evolves unitarily. In most of the studies on dynamics of open
systems, it is assumed that the open system and its environment are
at the initial moment of their joint evolution
factorized\cite{EBDavies,JPreskill}, that is they are described by
the density operator of the form
$$\rho_{sb}=\rho_{s}\otimes\rho_{b},$$
where $\rho_{s}$ is the initial state of system and $\rho_{b}$ is
the initial state of the environment. As the combined system is a
closed one, its evolution is unitary,
\begin{equation}
\rho_{sb}(t)=U_{sb}\rho_{sb}(0)U_{sb}^{\dagger},
\end{equation}
where $U_{sb}$ is the unitary operator.The interested system, as an
open one, than evolves in the following way
\begin{equation}\label{density matrix of system}
\rho_{s}(t)=tr_{b}\{U_{sb}\rho_{sb}U_{sb}^{\dagger}\}=tr_{b}\{\rho_{sb}(t)\}.
\end{equation}
The above equation is obtained by doing partial-trace on bath.
Kraus\cite{KKraus,K.Kraus} and Choi\cite{MChoi} showed that if the
above equation can be equivalently expressed in the form
\begin{equation}\label{ro system}
\rho_{s}(t)=\Phi(\rho_{s}(0))=\sum_{\mu\nu}M_{\mu\nu}(t)\rho_{s}(0)M_{\mu\nu}^{\dagger}(t),
\end{equation}
where the map $\Phi$ is described by
$\Phi:\rho_{s}(0)\rightarrow\rho_{s}(t)$ and $M_{\mu\nu}(t)$ satisfy
the following equation
\begin{equation}\label{Krasu}
\sum_{\mu\nu}M_{\mu\nu}^{\dagger}(t)M_{\mu\nu}(t)=I,
\end{equation}
it is said that evolution of $\rho_{s}(t)$ has the form of the Kraus
representation and the above equation is condition of CP
map\cite{JBouda}.

In this paper, we consider a model of a spin star configuration
which consists of N+1 localized spin-$\frac{1}{2}$ particles. One of
the spins is located at the center of the star, while the other
spins on concentric circles with different radii surrounding the
central spin layer by layer\cite{MMahdian}. The difference in radius
is because the coupling coefficients between layers spins and the
central spin is taken differently. At first, we obtain Kraus
representations and then arrive master equation with the use of
Kraus representations and show that they satisfy Eq.(4).

Maps are one of the main bases to obtain Kraus representations. In
addition, they are effective in obtaining exact NZ and TCL master
equation\cite{ASmirne}. NZ and TCL are used when the exact solution
of the master equation is a difficult task, therefore we can
approximate the master equation to solve\cite{HGrabert,HBreuer}.
These techniques were introduced by Nakajima (1958) and Zwanzig
(1960) and independently by the Brussels school (Prigogine,
1962)\cite{HBreuer}. They are widely used in non-equilibrium
statistical mechanics and have played an important role in various
fields of physics (e.g. in the theory of transport phenomena and
relaxation processes, in quantum theory damping, in super-radiance
and laser theory)\cite{RKühne}. Details of the application of this
method in the various fields and numerous references are given in
books and reviews articles of $Haken$\cite{HHaken,Haken},
$Agarwal$\cite{GSAgarwal} and $Haake$ \cite{FHaake}. By considering
the above model and exploiting the knowledge of the exact unitary
evolution, and the reduced dynamics, as well as a suitable matrix
representation of the dynamical maps, we can exhibit the exact TCL
and NZ equation of motion.

The paper is organized as follows. In Sec.(II), we will explain
about how to obtain map and Kraus representation and master
equation. Exact NZ and TCL master equation are discussed in
Sec.(III). Finally, in Sec.(IV), we consider a simple example that
shows vanishing quantum discord is not necessary for CP maps.

\section{Map, Kraus representation and Master equation}
\subsection{Background}
The first step in relating a general, non-Markovian master equation
$\dot{\rho}(t)=\Lambda[\rho(t)]$ to a corresponding CP map
$\rho(t)=\Phi_{t}[\rho(0)]$ relies on expressing this relationship
in matrix form. We need to write the matrix form to the basis of
$\{W_{a}\}$, that the basis denotes every convenient orthonormal
basis set for the Hermitian operators on the Hilbert space, i.e.
$$W_{a}^{\dagger}=W_{a},$$ $$tr[W_{a}W_{b}]=\delta_{ab}.$$
Now, according to the introduced basis, for any maps $\Phi$ (though
we will be concerned mainly with CP maps below) we can express both
$\rho$ and $\Phi(\rho)$ in matrix form as follows\cite{EAndersson},
\begin{equation}\label{map}
\Phi(\rho)=(Fr)^{T}W,
\end{equation}
where the evolution matrix $F$ is the following
\begin{equation}\label{26}
F_{kl}:=tr[W_{k}\Phi(W_{l})],
\end{equation}
and also vector $r$ is defined as
$$r_{l}:=tr[W_{l}\rho].$$
Indicating time-dependence of $\Phi$ and $F$ via $\Phi_{t}$ and
$F(t)$, and defining
\begin{equation}\label{24}
\rho(t):=\Phi_{t}[\rho(0)].
\end{equation}
Time evolution of $\rho$ can be expressed as
$$\dot{\rho}=[\dot{F}(t)r(0)]^{T}W.$$
Now, suppose that $\rho(t)$ satisfies a master equation of the form
\begin{equation}\label{master}
\dot{\rho}=\Lambda(\rho).
\end{equation}
Note that the linear map $\Lambda$ may be time-dependent. Now, this
master equation can be rewritten as
$$\dot{\rho}=[\dot{L}(t)r(t)]^{T}W.$$
Note that $L$ (and $F$) are necessarily real, but not symmetric in
general. Comparing the results of the last two paragraphs gives
\begin{equation}\label{L}
\dot{F}r(0)=Lr(t)=LFr(0).
\end{equation}
Hence, since by linearity this equation must hold for all vectors
$r(0)$ (whether or not they correspond to a density operator).

Choi has shown that complete positivity of $\Phi$ is equivalent to
the positivity of the particular matrix $S$ \cite{MDChoi,CMCaves},
and that the Kraus decompositions of $\Phi$ are related to the outer
product decomposition of $S$. The master equation gives us matrix
$L$, which in turn gives us a matrix $F$, characterising the linear
evolution map $\Phi_{t}$. In order to proceed from $F$ to a Kraus
decomposition, all we need to do is finding $S$ from $F$, and then
diagonalise it.  In fact, our construction of $S$ from $L$ allows us
to $determine$ whether or not, for a given proposed master equation,
the corresponding map is completely positive. Choi further
demonstrated that $\Phi$ is completely positive if and only if $S$
is positive, i.e., $S\geq0$ \cite{MDChoi,CMCaves}. The Hermitian
property implies that one can always decompose $S$ as a sum of outer
products, i.e.,
\begin{equation}\label{S}
S=\sum_{i}V_{i}(t)V_{i}^{\dagger}(t)
\end{equation}
for some set vectors$\{V(i)\}$ that we take $\{V(i)\}$ as follows
\begin{equation}\label{V}
V(i)=\sqrt{\mid\lambda_{i}\mid}\Pi(i),
\end{equation}
where $\Pi$ coefficients are eigenvectors of matrix $S$ and also the
Kraus operators are given by
\begin{equation}\label{kraus}
M_{i}:=\sum_{a}V(i)_{a}\beta_{a}.
\end{equation}
Noting that the $\beta_{a}$ form an orthonormal (non-Hermitian)
basis for the operators on the Hilbert space\cite{EAndersson}.

\subsection{Map, Master Equation and Kraus representation for spin star model for one-qubit system with layered environment in a channel}
We obtain all the issues described in the previous section for the
model which at first was mentioned. The Hamiltonian of the XX
central spin model that is composed by a localized spin, hereafter
called central spin, coupled to N spins of bath layers with the
different coupling constant $\alpha$, takes the form\cite{MMahdian}
\begin{equation}\label{intraction}
H=2(\sigma_{+}\Xi_{-}+\sigma_{-}\Xi_{+}),
\end{equation}
where $\Xi_{\pm}$ is denoted as
\begin{equation}\label{j positive}
\Xi_{+}=\sum_{\mu=1}^{n}\alpha_{\mu}J^{\mu}_{+},
\end{equation}
\begin{equation}\label{j negative}
\Xi_{-}=\sum_{\mu=1}^{n}\alpha_{\mu}J^{\mu}_{-},
\end{equation}
where $\sigma_{\pm}$ and $J^{\mu}_{\pm}$ are the Pauli operators
referring to the central spin and the surrounding spins respectively
, and $J^{\mu}_{\pm}$ is denoted as
\begin{equation}\label{Hamiltonian}
J^{\mu}_{\pm}\equiv\sum_{j=1}^{N_{\mu}}\sigma^{j}_{\pm}.
\end{equation}
So, evolution density matrix of system is as follows
\begin{equation}\label{density matrix of system(1)}
\rho_{S}(t)=\frac{1}{2}\left(
        \begin{array}{cc}
1+\upsilon_{3}(0)f_{3}&\upsilon_{-}(0)f_{12} \\
\upsilon_{+}(0)f_{12}&1-\upsilon_{3}(0)f_{3}\\
        \end{array}
        \right)
\end{equation}
where we have introduced the functions\cite{MMahdian}
\begin{equation}
f_{12}(t)\equiv
tr_{B}\{\cos[2th_{1}(\alpha_{1},\ldots,\alpha_{n})]\cos[2th_{2}(\alpha_{1},\ldots,\alpha_{n})]
\otimes 2^{-N}I_{B}\},
\end{equation}
and
\begin{equation}
f_{3}(t)\equiv
tr_{B}\{\cos[2th_{1}(\alpha_{1},\ldots,\alpha_{n})]\otimes
2^{-N}I_{B}\},
\end{equation}
where $h_{1}(\alpha_{1},\alpha_{2},\ldots,\alpha_{n})$ and
$h_{2}(\alpha_{1},\alpha_{2},\ldots,\alpha_{n})$ are
\begin{equation}
h_{1}=h_{1}(\alpha_{1},\alpha_{2},\ldots,\alpha_{n})=\sqrt{\sum_{\mu=1}^{n}\alpha^{2}_{\mu}J^{\mu}_{+}J^{\mu}_{-}},
\end{equation}
\begin{equation}
h_{2}=h_{2}(\alpha_{1},\alpha_{2},\ldots,\alpha_{n})=\sqrt{\sum_{\mu}^{n}\alpha^{2}_{\mu}J^{\mu}_{-}J^{\mu}_{+}}.
\end{equation}
To obtain the master equation, the matrix $F$ firs must be
determined, and checked as whether it is invertible. The basis
$\{W_{a}\}$ is chosen as $W_{a}=\frac{1}{\sqrt{2}}\sigma_{a}$, where
$\sigma_{a}$ are the Pauli operators. One finds
\begin{equation}\label{map(1)}
\Phi(W_{0})=\frac{1}{\sqrt{2}}\left(
        \begin{array}{cc}
1&0\\0&1\\
        \end{array}
        \right),
\end{equation}
\begin{equation}\label{map(2)}
\Phi(W_{1})=\frac{1}{\sqrt{2}}\left(
        \begin{array}{cc}
0&f_{12}\\f_{12}&0\\
        \end{array}
        \right),
\end{equation}
\begin{equation}\label{map(2)}
\Phi(W_{2})=\frac{1}{\sqrt{2}}\left(
        \begin{array}{cc}
0&-if_{12}\\if_{12}&0\\
        \end{array}
        \right),
\end{equation}
\begin{equation}\label{map(2)}
\Phi(W_{3})=\frac{1}{\sqrt{2}}\left(
        \begin{array}{cc}
f_{3}&0\\0&f_{3}\\
        \end{array}
        \right).
\end{equation}
The matrix $F$  follows via Eqs.(5) and (6) as
\begin{equation}\label{F(1)}
F=\left(
        \begin{array}{cccc}
1&0&0&0\\
0&f_{12}&0&0\\
0&0&f_{12}&0\\
0&0&0&f_{3}\\
        \end{array}
        \right),
\end{equation}

The solution for $L$ follows via Eq.(9)as
\begin{equation}\label{F(4)}
L=\dot{F}F^{-1}=\left(
        \begin{array}{cccc}
0&0&0&0\\
0&\frac{f'_{12}}{f_{12}}&0&0\\
0&0&\frac{f'_{12}}{f_{12}}&0\\
0&0&0&\frac{f'_{3}}{f_{3}}\\
        \end{array}
        \right).
\end{equation}
Now, to obtain the master equation in the form
$\dot{\rho}=\Lambda_{t}(\rho)$, we calculate the Choi matrix $R$ in
Ref.\cite{EAndersson},
\begin{equation}\label{R}
R=\left(
        \begin{array}{cccc}
\frac{f'_{3}}{2f_{3}}&\frac{f'_{12}}{f_{12}}&0&0\\
\frac{f'_{12}}{f_{12}}&\frac{f'_{3}}{2f_{3}}&0&0\\
0&0&-\frac{f'_{3}}{2f_{3}}&0\\
0&0&0&-\frac{f'_{3}}{2f_{3}}\\
        \end{array}
        \right),
\end{equation}
The master equation immediately follows via equation as
$$\Lambda_{t}(\rho):=\sum_{ab}R_{ab}(t)\beta_{a}\rho\beta_{b}^{\dagger},$$
\begin{equation}\label{master equation}
\dot{\rho}(t)=\Lambda_{t}[\rho(t)]=(\frac{f'_{3}}{2f_{3}}+\frac{f'_{12}}{f_{12}})\{\sigma_{+}\sigma_{-},\rho\sigma_{-}\sigma_{+}\}-\frac{f'_{3}}{2f_{3}}(\sigma_{+}\rho\sigma_{+}+\sigma_{-}\rho\sigma_{-}).
\end{equation}
Finally, we obtain Kraus representations via using of Eqs.(11) and
(12). At first, we calculate matrix $S$ as follows
\begin{equation}\label{S(1)}
S=\frac{1}{2}\left(
        \begin{array}{cccc}
1+2f_{12}+f_{3}&0&0&0\\
0&1-f_{3}&0&0\\
0&0&1-f_{3}&0\\
0&0&0&1-2f_{12}+f_{3}\\
        \end{array}
        \right).
\end{equation}
Now, we obtain Kraus representations
\begin{equation}\label{Krasu(1)}
M_{1}=\sqrt{\frac{1}{2}(1+2f_{12}+f_{3})}W_{0},
\end{equation}
\begin{equation}\label{Krasu(2)}
M_{2}=\sqrt{\frac{1}{2}(1-f_{3})}W_{1},
\end{equation}
\begin{equation}\label{Krasu(3)}
M_{3}=\sqrt{\frac{1}{2}(1-f_{3})}W_{2},
\end{equation}
\begin{equation}\label{Krasu(4)}
M_{4}=\sqrt{\frac{1}{2}(1-2f_{12}+f_{3})}W_{3},
\end{equation}
that they satisfy Eq.(4).
\section{exact Nakajima-Zwanzig and time-convolutionless master equation}
\subsection{Background}

With the aid of the knowledge of the exact time evolution, and using
the representation of maps in terms of matrices, we now explicitly
obtain two kinds of exact equations of motion for the reduced
system's dynamics. We first consider a master equation in
differential form with a generator local in time, that is, the TCL
master equation. Assuming the existence of such a generator
$K_{TCL}(t)$, it should obey the equation\cite{ASmirne}
\begin{equation}\label{TCL}
\dot{\rho}(t)=K_{TCL}(t)\rho(t),
\end{equation}
which, due to $\rho(t)=\Phi(t)\rho(0)$, is satisfied upon
identifying
\begin{equation}\label{K(tcl)}
K_{TCL}=\dot{\Phi}(t)\Phi^{-1}(t)
\end{equation}
or, in terms of matrices,
\begin{equation}\label{matrix(K)}
P_{TCL}=\dot{F}(t)F^{-1}(t),
\end{equation}
this expression holds when $F$ is invertible.

Now, we can obtain the NZ master equation according to the TCL
master equation when memory kernel exists. In this case the memory
kernel $K_{NZ}(t)$ should obey the convolution equation
\begin{equation}\label{K(NZ)}
\dot{\rho}(t)=(K_{NZ}\circ\rho)(t),
\end{equation}
so that in view of Eq.(7), one has the relation
\begin{equation}\label{hat(K(NZ))}
\widehat{K}_{NZ}(u)=u\mathbb{I}-\widehat{\Phi}^{-1}(u),
\end{equation}
where the hat denotes the Laplace transform, and therefore in matrix
representation,
\begin{equation}\label{hat(J(NZ))}
\widehat{P}_{NZ}(u)=u\mathbb{I}-\widehat{F}^{-1}(u).
\end{equation}
\subsection{Nakajima-Zwanzig($NZ$) and Time-Convolutionless($TCL$) for spin star model for one-qubit system with layered environment in a channel}
The technique used in the previous subsection to obtain the TCL and
NZ equation of motion for a model whose evolution is known, by
exploiting the representation of maps in terms of matrices, is
applicable for a detailed study of Model in Sec.(IIB).

Now, starting from Eq.(17) and exploiting the same strategy used in
Sec.(IIIA), one immediately obtains, for the matrix representation
of the TCL generator, the expression
\begin{equation}\label{1}
P_{TCL}(t)=\left(
        \begin{array}{cccc}
0&0&0&0\\
0&\gamma_{1}&0&0\\
0&0&\gamma_{1}&0\\
0&0&0&\gamma_{2}\\
        \end{array}
        \right),
\end{equation}
where $\gamma_{1}$ and $\gamma_{2}$ are
$$\gamma_{1}=-2(h_{1}\tan[2h_{1}t]+h_{2}\tan[2h_{2}t]), \gamma_{2}=-4h_{1}\tan[4h_{1}t].$$

And one can also determine the expression of the NZ memory kernel,
whose Laplace transform is given by
\begin{equation}\label{2}
\widehat{P}_{NZ}(u)=\left(
        \begin{array}{cccc}
0&0&0&0\\
0&\eta_{1}&0&0\\
0&0&\eta_{1}&0\\
0&0&0&\eta_{2}\\
        \end{array}
        \right),
\end{equation}
where $\eta_{1}$ and $\eta_{2}$ are
$$\eta_{1}=u-\frac{2^{N}[u^{4}+8(h_{1}^{2}+h_{2}^{2})u^{2}+16(h_{1}^{2}-h_{2}^{2})^{2}]}{u^{3}+4(h_{1}^{2}+h_{2}^{2})u}, \eta_{2}=u-\frac{2^{N}[u^{2}+16h_{1}^{2}]}{u}.$$

So the master equation in operator form reads\cite{ASmirne},
\begin{equation}\label{3}
k_{TCL}(t)\rho=-\frac{1}{2}\eta_{2}(\sigma_{+}\rho\sigma_{-}-\frac{1}{2}\{\sigma_{-}\sigma_{+},\rho\})-\frac{1}{2}\eta_{2}(\sigma_{-}\rho\sigma_{+}-\frac{1}{2}\{\sigma_{+}\sigma_{-},\rho\})+\frac{1}{4}(\eta_{2}-2\eta_{1})(\sigma_{z}\rho\sigma_{z}-\rho),
\end{equation}
and
\begin{equation}\label{4}
\widehat{K}_{NZ}(u)\rho=-\frac{1}{2}\gamma_{2}(\sigma_{+}\rho\sigma_{-}-\frac{1}{2}\{\sigma_{-}\sigma_{+},\rho\})-\frac{1}{2}\gamma_{2}(\sigma_{-}\rho\sigma_{+}-\frac{1}{2}\{\sigma_{+}\sigma_{-},\rho\})+\frac{1}{4}(\gamma_{2}-2\gamma_{1})(\sigma_{z}\rho\sigma_{z}-\rho).
\end{equation}
It immediately appears that, given the time evolution map from the
relation Eqs.(41) and (42), one can directly obtain the generator of
the master equation in TCL form or the memory kernel for the NZ
form, respectively, without resorting to the evolution of the whole
perturbative series.Obviously, given the exact time evolution, one
dose not need the equations of motion.

\section{Initial quantum correlation and CP maps}
\subsection{Background}
Recently, a relation between CP maps and quantum discord has been
put forth \cite{CARodrigues,lidar}. quantum discord, first
introduced by Ollivier and Zurek \cite{HOllivier}, that captures the
difference of two natural quantum extensions of the classical mutual
information, can be used as a measure of the quantum correlations.
Although quantum discord is equal to the entanglement for pure
states, it includes the quantum correlation which are contained in
mixed states that are not entangled. In \cite{CARodrigues}, it is
shown that if the initial system-bath state has vanishing quantum
discord, then the dynamics of system can be described by a CP map.
In \cite{lidar} Shabani and Lidar made a strong claim: $\textit{The
reduced dynamics of a system is completely-positive,}$ $\textit{for
any coupling with the bath, only if the initial}$
 $\textit{system-bath
state has vanishing quantum discord}$ $\textit{as measured by the
system.}$

Here we show that underlying assumption in Ref.\cite{lidar} limit
the generality of the constructed reduced dynamics. To do this we
consider two different paradigms to ask the question: given any
unitary evolution for the system-bath, can we define a CP map taking
a family of initial states $\{\rho^{S}(0)\}$ to final states
$\{\rho^{S}(t)\}$?
\subsection{Vanishing quantum discord is not necessary for CP maps}
 Now, we prove our claim that vanishing quantum discord is not necessary for CP with notice to
below example. In this example, first we obtain Kraus representation
with map and Eq.(3) and then we prove that map is CP with notice to
Eq.(4). On the other hand, we compute quantum discord and it shows
that we can choose conditions for nonzero quantum discord.

We choose the same model as that Ref.\cite{HHayashi} has used. That
is to consider a combined system composed of two spin-1/2 subsystems
with the interaction Hamiltonian
\begin{equation}\label{21}
H_{sb}=\sigma_{x}\otimes\frac{1}{2}(I-\sigma_{z})+I\otimes\frac{1}{2}(I+\sigma_{z}),
\end{equation}
where $\sigma_{x}$ and $\sigma_{z}$ are Pauli spin operators. In
this model, the first qubit plays the role of the open system while
the second qubit plays the role of the environment. The interaction
described by the Hamiltonian corresponds to the well-known
controlled- NOT gate \cite{JPreskill,PStelmachovic}. The unitary
evolution operator is given by $U_{sb}(t)=e^{-iH_{sb}t}$, explicitly
\begin{equation}\label{22}
U_{sb}(t)=\left(
        \begin{array}{cccc}
e^{-it}&0&0&0\\
0&\cos{t}&0&-i\sin{t}\\
0&0&e^{-it}&0\\
0&-i\sin{t}&0&\cos{t}\\
        \end{array}
        \right),
\end{equation}

 So, first we obtain $\rho_{sb}(0)$ as follows
\begin{equation}\label{11}
\rho_{sb}(0)=\frac{1}{4}(I^{s}\otimes
I^{b}+\sum_{i=1}^{3}c_{i}\sigma^{s}_{i}\otimes\sigma^{b}_{i}),
\end{equation}
and also we can obtain $\rho_{s}(0)$ with trace of $\rho_{sb}(0)$ as
follows
\begin{equation}\label{12}
\rho_{s}(0)=\frac{1}{2}I,
\end{equation}
where $\sigma_{i}s$ are Pauli matrices. From Eqs. (2), (10) and
(11), we get the density matrix of the system
\begin{equation}\label{23}
\rho_{s}(t)=\frac{1}{4}\left(
        \begin{array}{cc}
2+c_{3}(1-\cos{(2t)})&-i c_{3}\sin{(2t)}\\i c_{3}\sin{(2t)}&2-c_{3}(1-\cos{(2t)})\\
        \end{array}
        \right).
\end{equation}
So, maps can be written, with notice to Eqs (7) and (49), as
$$\Phi(W_{0})=\frac{1}{2\sqrt{2}}\left(
        \begin{array}{cc}
2+c_{3}(1-\cos{(2t)})&-i c_{3}\sin{(2t)}\\i c_{3}\sin{(2t)}&2-c_{3}(1-\cos{(2t)})\\
        \end{array}
        \right),$$
\begin{equation}\label{13}
\Phi(W_{1})=\Phi(W_{2})=\Phi(W_{3})=0.
\end{equation}
Now, we can obtain matrix of F and then matrix of S by having maps.
On the other hand, we can arrive Kraus representations with have
matrix of S as follows
\begin{equation}\label{14}
M_{1}=\frac{1}{2}\Gamma(t)\sin{(t)}W_{0}+\frac{i}{2}\Gamma(t)\cos{(t)}W_{1}-\frac{\sqrt{2}}{4}W_{3},
\end{equation}
\begin{equation}\label{15}
M_{2}=-\frac{1}{2}\Gamma(t)\cos{(t)}W_{0}+\frac{i}{2}\Gamma(t)\sin{(t)}W_{1}+\frac{\sqrt{2}}{4}W_{2},
\end{equation}
\begin{equation}\label{16}
M_{3}=\frac{1}{2}\Gamma(t)\sin{(t)}W_{0}+\frac{i}{2}\Gamma(t)\cos{(t)}W_{1}+\frac{\sqrt{2}}{4}W_{3},
\end{equation}
\begin{equation}\label{17}
M_{4}=\frac{1}{2}\Gamma(t)\cos{(t)}W_{0}-\frac{i}{2}\Gamma(t)\sin{(t)}W_{1}+\frac{\sqrt{2}}{4}W_{2},
\end{equation}
where $\Gamma(t)=\sqrt{1-c_{3}\sin{(t)}}$ and Kraus operators
$\{M_{i}\}$ satisfy Eq.(4) which expresses that our choose map is
CP.

However, given the quantum discord calculated in Ref.\cite{SLuo} and
Eq.(4), we can conclude that according to Kraus representations
obtained, Eq.(4) will always be satisfied regardless of the values
of $c_{3}$ and the values of quantum discord and it is clearly
violation of claim of Ref.\cite{lidar}. So, we understand that
vanishing quantum discordC is not necessary for CP map.
\section{conclusion}
In this paper, we have started from obtained maps of spin star model
for one-qubit system with layered environment and then by use of
these maps, computed the matrix $F$. With the matrix $F$, first we
have obtained the Choi matrix $S$ and then by use of matrix $S$,
have arrived Kraus representation and master equation that clearly,
we understood that Kraus representation satisfy Eq.(4). It expresses
that maps are CP. On the other hand, by use of the matrix $F$ we
have arrived matrix representations of the $TCL$ and $NZ$ generator
and then with use of matrix representations we have obtained
operator representations of the $TCL$ and $NZ$ generator. In the
end, with the two subjects expressed (satisfies the equation (4) and
the calculated quantum discord), we have proven that vanishing
quantum discord is not necessary for CP.


\newpage

\begin{thebibliography}{0}


\bibitem{MKeyl} M. Keyl and R.F. Werner, How to Correct Small Quantum Errors,
Lecture Notes in Physics Volume 611, Springer, (2002).
\bibitem{LGyongyosi} L. Gyongyosi and S.Imre, Properties of the Quantum
Channel, (2012).
\bibitem{HAnsari}H. Ansari, A. Parameswaran, L. Antani, B. Aditya, A. Taly and L.
Kumar, Quantum Cryptography and Quantum Computation (Department of
Computer Science and Engineering Indian Institute of Technology,
Bombay Mumbai), eprint: quant-ph/1208.1270v5(2003)
\bibitem{IDevetak} I. Devetak and P. W.Shor, The Capacity of a Quantum Channel for Simultaneous Transmission of Classical and Quantum Information, Commun. Math. Phys. 256 (2004).
\bibitem{PWShor} P. W.Shor, Equivalence of additivity questions in quantum information theory, Commun. Math. Phys. 246 (2004).
\bibitem{RVasile} R. Vasile, S. Maniscalco, M. G. A. Paris, H.-P. Breuer, J. Piilo, Quantifying non-Markovianity of continuous-variable Gaussian dynamical maps
 ,Phys.
Rev. A 84, 052118 (2011).
\bibitem{HPBreuer} H.-P. Breuer et al. , Measure for the
Degree of Non-Markovian Behavior of Quantum Pro- cesses in Open
Systems ,Phys. Rev. Lett. 103, 210401 (2009).
\bibitem{HBreuer} H. Breuer and F. Petruccione, The theory of open quantum systems
(Oxford University Press, 2007).
\bibitem{HCarmichael} H. Carmichael, An Open Systems Approach to Quantum Optics,
Lecture Notes in Physics, Vol. 18 (Springer-Verlag, Berlin, 1993).
\bibitem{CSlichter} C. Slichter, Principles of Magnetic Resonance, Springer Series
in Solid-State Sciences, Vol. 1 (Springer, Berlin, 1996).
\bibitem{RAlicki} R. Alicki, D. A. Lidar, and P. Zanardi, Internal consistency of fault-tolerant quantum error correction in light of rigorous derivations of the quantum Markovian limit, Phys. Rev. A 73, 052311
(2006).
\bibitem{UWeiss} U. Weiss, Quantum Dissipative Systems (World Scientific,
Singapore, 1993).
\bibitem{PStelmachovic} P. Stelmachovic and V. Buzek, Dynamics of open quantum systems initially entangled with environment: Beyond the Kraus representation , Phys. Rev. A 64, 062106 (2001); 67, 029902(E)(2001).
\bibitem{DSalgado} D. Salgado, and J. L. S´anchez-G´omez, Comment on Dynamics of open quantum systems initially entangled with environment: Beyond the Kraus representation ,eprint: quant-ph/0211164(2002).
\bibitem{HHayashi} H. Hayashi, G. Kimura and Y. Ota, Kraus representation in the presence of initial correlations , Phys. Rev. A 67, 062109 (2003).
\bibitem{DMTong} D. M. Tong, Jing-Ling Chen, L. C. Kwek and C. H. Oh, Kraus representation for density operator of arbitrary open qubit system ,eprint: quant-ph/0311091(2003).
\bibitem{NArshed} N. Arshed, A. H. Toor and D. A. Lidar, Channel capacities of an exactly solvable spin-star system , Phys. Rev. A. 81, 062353 (2010).
\bibitem{KMFRomero} K. M. F. Romero and R. L. Franco, Simple non-Markovian microscopic models for the depolarizing channel of a single qubit , Phys. Scr. 86, 065004 (9pp)
(2012).
\bibitem{MANielsen} M. A. Nielsen and I. L. Chuang, QuantumComputation and Quantum
Information (Cambridge University Press, 2000).
\bibitem{KKraus} K. Kraus, States, Effects and Operations (Spring-Verlag,
Berlin, 1983).
\bibitem{EBDavies} E.B. Davies, Quantum Theory of Open Systems (Academic,
London, 1976).
\bibitem{JPreskill} J. Preskill, Lecture notes:Information for
Physics 219/Computer Science 219, Quantum
Computation,www.theory.caltech.edu/people/preskill/ph229. 5.
\bibitem{K.Kraus} K. Kraus, "General state changes in quantum theory" Ann. Physics 64, 311–335
(1971).
\bibitem{MChoi} M. Choi, Completely Positive Linear Maps on Complex matrices,
Linear Algebra and Its Applications, 285–290, (1975).
\bibitem{JBouda} J. Bouda and V. Bu¢zek, Purification and correlated measurements of bipartite mixed states , Phys. Rev. A 65, 034304 (2003).
\bibitem{MMahdian} M. Mahdian and H. Mehrabpour, Exact dynamics of one-qubit system in  layered  environment, eprint: quant-ph/0701483(2013).
\bibitem{ASmirne} A. Smirne and B. Vacchini, Nakajima-Zwanzig versus time-convolutionless master equation for the non-Markovian dynamics of a two-level system , Phys. Rev. A.82, 022110 (2010).
\bibitem{HGrabert} H. Grabert, Projection operator techniques in nonequilibrium statistical mechanics, Springer Tracts in Modern Physics, Band 95,
(1982).
\bibitem{RKühne} R. Kühne, P. Reineker, Zeitschrift für Physik B Condensed Matter,
Springer (1978).
\bibitem{HHaken} H. Haken, Laser Theory, (Encyclopedia of Physics XXV/2c, ed. Flügge, S. Berlin-Heidelberg-New York: Springer, 1970).
\bibitem{Haken} H. Haken, Cooperative phenomena in systems far from thermal equilibrium and in nonphysical systems , Rev. Mod. Physics 47, 67 (1975).
\bibitem{GSAgarwal} G. S. Agarwal, Master Equation Methods in Quantum Optics, (Progress in Optics XI, ed. Wolf, E. Amsterdam-London: North-Holland Publishing Company
1973).
\bibitem{FHaake} F. Haake, Statistical Treatment of Open Systems by Generalized Master Equations, (Springer Tracts in Modern Physics, Volume 66. Berlin-Heidelberg-New York: Springer
1973).
\bibitem{EAndersson} E. Andersson, J.D. Cresser, M.J.W. Hall, Finding the Kraus decomposition from a master equation and vice versa , Journal of Modern Optics,
(2007).
\bibitem{MDChoi} M. -D. Choi, Completely positive maps on complex matrices , Lin. Alg. Appl. 10 285 (1975).
\bibitem{CMCaves} C. M. Caves, Quantum error correction and reversible operations , J. Supercond. 12(6) 707 (1999); also published as quant-ph/9811082.
\bibitem{CARodrigues} C. A. Rodrigues-Rosario, K. Modi, A. Kuah, A. Shaji, and E. C. G. Sudarshan, Completely positive maps and classical correlations , J. Phys. A: Math. Gen. 41,
205301 (2008).
\bibitem{lidar} A. Shabani and D. Lidar, Vanishing quantum discord is necessary and sufficient for completely positive maps , Phys. Rev. Lett. 102,
100402 (2009).
\bibitem{HOllivier} H. Ollivier and W. H. Zurek, Quantum Discord: A Measure of the Quantumness of Correlations , Phys. Rev. Lett. 88, 017901 (2001).
\bibitem{SLuo} S. Luo, Quantum discord for two-qubit systems , Phys. Rev. A. 77, 042303 (2008).
\end{thebibliography}
\end{document}